\documentclass[twocolumn]{article}
\usepackage[T1]{fontenc}
\usepackage{kpfonts}

\usepackage{xurl}
\usepackage{booktabs}

\usepackage{geometry}
\usepackage[square, numbers]{natbib}
\usepackage{xcolor}
\definecolor{unicolor}{HTML}{990000}
\usepackage[hidelinks]{hyperref}
\hypersetup{
    colorlinks=true,
    linkcolor=blue, 
    filecolor=black,
    citecolor=unicolor,
    urlcolor=blue
    }

\usepackage{graphicx}

\usepackage{amssymb}
\usepackage{amsmath}

\usepackage{authblk}
\title{\vspace{-1em}How the cascade inference problem\\ distorts information diffusion}

\author[a,b,$\dagger$,$\diamond$]{Matthew R. DeVerna}
\author[b,c,$\dagger$]{Francesco Pierri}
\author[c]{Rachith Aiyappa}
\author[c,d]{Diogo Pacheco}
\author[c]{John Bryden}
\author[c]{Filippo Menczer}

\affil[a]{\small Tech Impact and Policy Center, Stanford University, Stanford, California, 94305}
\affil[b]{\small Observatory on Social Media, Indiana University Bloomington, Indiana, 47408}
\affil[c]{\small Department of Electronics, Information and Bioengineering, Politecnico di Milano, Italia, 20133}
\affil[d]{\small Department of Computer Science, University of Exeter, Exeter, United Kingdom, EX4 4RN}
\affil[$\dagger$]{\small Contributed equally.}
\affil[$\diamond$]{\small Correspondence: \texttt{mdeverna@stanford.edu}.}

\date{}
\begin{document}

\maketitle

\begin{abstract}
    To analyze the flow of information online, experts often rely on platform-provided data from social media companies, which typically attribute all resharing actions to an original poster. 
    This obscures the true dynamics of how information spreads online, as users can be exposed to content in various ways. 
    While most researchers analyze data as it is provided by the platform and overlook this issue, some attempt to infer the structure of  information cascades.
    However, the absence of ground truth about actual diffusion cascades makes it impossible to verify the efficacy of these efforts.
    We propose a novel parametric reconstruction approach and use it to investigate how overlooking cascade reconstruction distorts analyses of social influence, community detection, and information diffusion. 
    Two case studies involving data from Twitter and Bluesky reveal that cascade inference significantly impacts the identification of both influential users and communities, therefore affecting downstream analyses in general. 
    Analysis of the diffusion of over 40,000 true and false news stories on Twitter reveals that the assumptions made during the reconstruction procedure drastically distort both microscopic and macroscopic properties of cascade networks. 
    This work highlights the challenges of studying information spreading processes on complex networks and has significant implications for the broader study of digital platforms.
\end{abstract}

\section*{Introduction}
\label{sec:intro}

The digital age has woven technology into nearly every aspect of daily life, generating an unprecedented volume of data on human behavior and societal trends~\cite{Levin2021Feb, sagiroglu2013big, Hilbert2011Feb}. 
This abundance of data has catalyzed the rise of computational social science, a field that leverages computational techniques to analyze and interpret digital trace data, providing novel insights into human behavior and society at large~\cite{Hofman2021Jul, Lazer2020Aug, Salganik2018, Lazer2009Feb, Watts2007Feb}.
Over the past two decades, social media platforms have become a primary source of data, typically accessed through Application Programming Interfaces (APIs) available to the public~\cite{jurgens2016tutorial}. 
This access has been vital not only for academic research but also for government and industry sectors, highlighting the pivotal role of digital data in modern research landscapes~\cite{Meyer2023Aug, Davidson2023Dec}. 

The importance of this data is reflected in the rapid growth of the scientific literature on the diffusion of online information, which has surged dramatically over the past decade. 
A bibliographic analysis (see the Appendix) shows that between 2018--2023, over a thousand articles from the Social Sciences, Physics, Engineering, Medicine, and other fields have been published annually on this topic. 
This vast literature has influenced our understanding of critical societal challenges related to public health~\cite{Tsao2021Mar, Schillinger2020Aug, Kass-Hout2013Dec, Dredze2012Aug}, political communication~\cite{persily2020social, Tucker2018Mar, Lazer2018Mar, Stieglitz2013Dec}, disaster response~\cite{Reuter2018Apr, Houston2015Jan, Alexander2014Sep}, collective action~\cite{Steinert-Threlkeld2017May, Gonzalez-Bailon2016Jan, Segerberg2011Jul}, and human attention~\cite{Lazer2020Jan, lorenz2019accelerating}, demonstrating the interdisciplinary appeal and broad impact of studying online information diffusion.

\begin{figure*}[t]
    \centering
    \includegraphics[width=.8\linewidth]{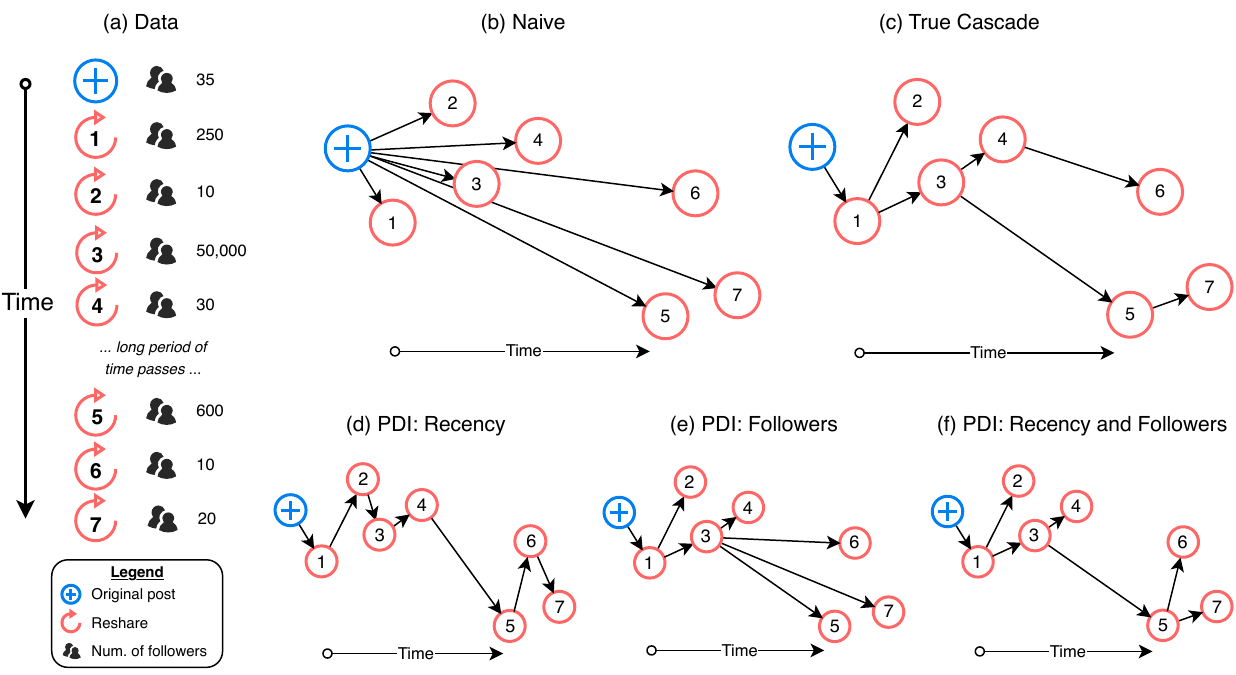}
    \caption{Cascade reconstruction with Probabilistic Diffusion Inference.
    Panels (a–c) illustrate the cascade inference problem.
    (a):~Hypothetical empirical data of a message cascade with an original post (blue cross) and a sequence of resharing actions (red circles) over time.
    Each post is associated with a timestamp (represented by the time sequence) and the number of followers of the resharing user (next to the user icon).    
    (b):~The naive cascade constructed from platform-provided data, which assumes that every user directly reshared the original post. 
    (c):~The true cascade, reflecting the actual parent-child relationships.
    Panels (d, e, f) demonstrate different cascade reconstructions when applying various PDI assumptions.
    The recency assumption (d) prioritizes users who reshared the content more recently, capturing temporal dynamics.
    The followers assumption (e) gives higher resharing likelihood to users with more followers, emphasizing popularity.
    Incorporating both assumptions (f) captures both temporal activity and popularity into the cascade reconstruction.
    }
    \label{fig:reconstruction}
\end{figure*}

Despite significant advances in the field, data from social media platforms present important limitations~\cite{Lazer2020Aug, Lazer2020Jan}.
For example, dynamic socio-technical systems continuously change, influencing user behavior in non-transparent ways even as we attempt to study them~\cite{Lazer2021Jul}. 
The influence of evolving and opaque platform algorithms adds further complexity to analyses~\cite{Wagner2021Jul}. 
Here we focus on an often-overlooked issue arising from the difficulty to reconstruct information sharing cascades. 
We refer to this challenge as the \textit{Cascade Inference Problem}: to recover the true structure of a social media post diffusion tree, where a user is connected to another from whom they were exposed to a post they reshared.
This challenge arises because we cannot fully observe who exposed a post to whom. 
Some platforms, like Facebook and Instagram, only provide aggregated data about cascade sizes. 
Others, like Twitter (now X), Mastodon, Bluesky, and Threads, provide more data but typically obscure the pathways of information diffusion by attributing all resharing actions to the original poster. 
This misrepresentation conceals the true dynamics of how information spreads, hindering our ability to fully understand these processes (cf. Figure~\ref{fig:reconstruction}~(b)~vs.~(c)).
To address this problem, researchers have proposed methods for inferring the structure of information cascades to predict the actual patterns of online information flow~\cite{Bakshy2011influencer, Gomez-Rodriguez-2012, Cogan2012Aug, Dow_Adamic_Friggeri_2013, Friggeri_Adamic_Eckles_Cheng_2014, Taxidou2014Apr, DeNies2015Oct, goel2016structural, Vosoughi2017Jul, vosoughi2018spread, Cinelli2022Sep}.
However, the problem is challenging because the reconstruction process must rely on numerous assumptions, introducing significant potential for systematic error. 
Such potential for error rises as cascades grow: with more nodes, accurately identifying a new post's parent becomes increasingly difficult~\cite{Bollenbacher2021Dec}. 
In rare cases, researchers may collaborate with platforms, which have access to detailed information about sharing actions~\cite{Gonzalez-Bailon2024Dec, bakshy2012role, Dow_Adamic_Friggeri_2013, Friggeri_Adamic_Eckles_Cheng_2014}. 
Yet even these datasets may contain errors, and the general lack of ground-truth data renders validation of reconstruction techniques impossible~\cite{Davidson2023Dec}. 

\subsection*{Cascade reconstruction problems}
\label{sec:relatedwork}

Researchers have long sought to reconstruct how information spreads online~\cite{Bakshy2011influencer, Gomez-Rodriguez-2012, Cogan2012Aug, Dow_Adamic_Friggeri_2013, Friggeri_Adamic_Eckles_Cheng_2014, Taxidou2014Apr, DeNies2015Oct, goel2016structural, Vosoughi2017Jul, vosoughi2018spread, Cinelli2022Sep}. 
Gómez-Rodríguez et al.~\cite{Gomez-Rodriguez-2012} framed this challenge as a network-inference problem, estimating diffusion pathways across blogs and news sites. 
Within the social media context, Cogan et al.~\cite{Cogan2012Aug} introduced an early method to rebuild reply-based conversation trees on Twitter.
Because direct replies explicitly identify whom a user is responding to, this task is considerably simpler than reconstructing reshare cascades.
Other efforts modeled how message content evolves over time~\cite{DeNies2015Oct} and across communities~\cite{Hu2015CommunityDiffusion}, while separate work aimed to predict the eventual size of cascades~\cite{Cheng2014cascadepredict, WengICWSM14}.
Such studies advanced theoretical understanding but did not recover the exact user-to-user chains underlying real information diffusion.

Goel et al.~\cite{goel2016structural} took a key step by inferring Twitter retweet cascades from follower-graph data---leveraging who follows whom to estimate likely exposure.
Vosoughi et al.~\cite{vosoughi2018spread} later popularized a similar ``Time-Inferred Diffusion'' (TID) approach to analyze a large set of true and false rumor cascades, reporting that falsehoods spread farther than truths.
Given the challenge of the reconstruction task, the data from this study was subsequently reanalyzed in many other domains~\cite{Bollenbacher2021Dec,Prollochs2023Oct,Naumzik2022Apr,Prollochs2021Nov,Juul2021Nov,Rosenfeld2020Apr,Ducci2020Aug}.

Industry researchers approached the cascade inference problem using richer internal company data.
At Facebook, for example, ``rechaining'' methods linked reshares based on user click behavior---if user C reshared A's post after first clicking B's reshare, the inferred edge connected C to B rather than A.
This approach was applied to rumor dynamics~\cite{Friggeri_Adamic_Eckles_Cheng_2014} and the role of social networks in information diffusion~\cite{bakshy2012role}.
In a recent collaboration with Meta, González-Bailón et al.~\cite{Gonzalez-Bailon2024Dec} used internal platform data to reconstruct roughly one billion Facebook resharing trees by relying exclusively on who clicked whose reshare button.
Similar approaches have been applied to analyses of Weibo data~\cite{Cao2017deephawkes, Chen2019InfoDiffPred}.
Applying this approach to the earlier example, an edge would be drawn from C to A, ignoring C's exposure to B's content. 

This body of work highlights how seemingly small reconstruction decisions can reshape the inferred structure of information cascades and, in turn, our understanding of diffusion at scale.
Yet all existing approaches depend on simplifying assumptions about exposure, timing, or user behavior that constrain the recovery of diffusion pathways, giving rise to the cascade inference problem.

Even with perfect click or view data, it is often unclear how to reconstruct a cascade.
For example, in the earlier ``rechaining'' example we might construct the cascade A~$\rightarrow$~B~$\rightarrow$~C after observing C reshared A's post only after viewing B's reshare.
Yet, as the Facebook researchers themselves noted, 
\begin{quote}
    ``This process is naturally not without error. For example if C clicks on multiple reshares, it is unclear which one to attribute to.''~\cite[pg. 147]{Dow_Adamic_Friggeri_2013}
\end{quote}
Moreover, researcher-defined assumptions persist when users do not click on multiple reshares.
Even in this simple case, we do not know whether to link C to A, whose reshare button C \textit{clicked}, or to B, whose reshare C \textit{viewed} before resharing.
Which edge, then, represents the ``true'' pathway of diffusion? 
González-Bailón et al.~\cite{Gonzalez-Bailon2024Dec} avoid this ambiguity by drawing edges solely based on which reshare button a user clicks.
The authors note that this yields ``very accurate representations of resharing behavior'' but not ``exact representations of all the causal pathways that allow people to become aware of content''~\cite[pg. S25]{Gonzalez-Bailon2024Dec}. 
While this distinction is valid, much of the diffusion literature is explicitly concerned with causal mechanisms, such as influence~\cite{Bakshy2011influencer,Dow_Adamic_Friggeri_2013,bakshy2012role}. 
Moreover, such an approach typically requires collaboration with major social media companies.

Relying on limited platform-provided API data introduces additional challenges.
Follower-graph-based approaches, such as TID, assume that users reshare content only from accounts they follow---a premise that is invalidated by the many ways users encounter content from non-followed accounts: algorithmic feeds, search, word-of-mouth, and so on.
Today, 50\% or more of X's algorithmically driven ``For You'' feed consists of content from accounts that users do not follow, and only 7\% and 17\% of time spent on Instagram and Facebook, respectively, is devoted to content from one's ``friends''~\cite{twitterAlgo, ftc2023meta}.
Even if the follow-only assumption held, additional assumptions are still required to handle the common case in which a user follows multiple candidate sources and one must be selected for attribution.

\subsection*{Research overview}
\label{sec:overview}

Despite growing recognition of these limitations, most studies continue to analyze platform-provided data without reconstructing cascades---introducing unknown biases into analyses.
These distortions have implications for addressing societal challenges such as the spread of misinformation, online polarization, and the fair identification of credible voices in public discourse. 
Therefore, we investigate the following research question: \textit{To what extent do cascade reconstruction choices distort our understanding of online information diffusion}?

To address this question, we first quantify how bypassing the cascade reconstruction process altogether impacts measures of social influence and community detection in two case studies on Twitter and Bluesky. 
After determining that this omission dramatically affects assessments of both node influence and the identification of communities, we investigate the structural effects of different reconstruction approaches.
Leveraging a widely studied dataset of over 40,000 Twitter news cascades~\cite{vosoughi2018spread}, we uncover substantial discrepancies in cascades at both micro and macro levels. 

Our findings indicate that transparent modeling of online information diffusion is essential to avoid misleading conclusions about influence and reach on digital platforms, ensuring that such analyses support rather than distort public understanding of online ecosystems.

\section*{Results}

\subsection*{Cascade reconstruction}

To understand how reconstructing information cascades impacts various analyses, we first introduce a general, parametric method that infers information (or message) cascades on microblogging platforms by leveraging empirical data about resharing activities.
A \textit{message cascade} is a tree structure where the root is the original poster of the message and a parent node's children are the users who reshared the message because they saw the parent's post. 
For each node in the cascade, the method infers the \textit{parent} node, i.e., the prior node within the tree (user who previously posted or reposted the same message) that led to the resharing action.  
Linking all the posters of a message through these parent-child connections forms the message cascade. 

Our method, called Probabilistic Diffusion Inference (PDI), relies on assumed probability distributions to weigh the likelihood of potential parents being the true parent within an information cascade. 
While this approach can flexibly incorporate any researcher-formulated probability distribution to capture the latest knowledge or potential platform changes, we adopt two assumptions based on previous work~\cite{vosoughi2018spread} about which users are more likely to be the parent of a resharer: users with more followers (\textit{followers} assumption) and users who are more recently active in the cascade (\textit{recency} assumption).
These assumptions are visually represented in Figure~\ref{fig:reconstruction}(d, e, f).

To model these assumptions, we calculate two probabilities for each potential parent node: one based on their number of followers and the other taking into account the recency of their activity.
A parameter $\alpha$ controls how much emphasis is placed on recency, with higher values giving more importance to recent posts.
The relative influence of these two factors is adjusted using a parameter $\gamma$---higher values give more weight to follower counts, while lower values prioritize reshare recency.
Further details on PDI and these assumptions can be found in the Materials and Methods. 

A set of cascade trees reconstructed from the data with the PDI method can be combined into a weighted \textit{resharing network}. 
Nodes in this network represent users and edges capture the flow of information.
Specifically, a link $(i \rightarrow j, w)$ represents a directed edge from user $i$ to user $j$, weighted by $w$, the number of times user $j$ reshared user $i$'s content. 
However, unlike reconstruction methods that generate a single cascade in deterministic fashion~\cite{vosoughi2018spread, goel2016structural}, PDI can stochastically generate many different realizations of each cascade.
This allows us to construct many versions of the weighted resharing network. 

\subsection*{Social influence measurement}

Pinpointing the most influential individuals within social networks is a critical and widely studied challenge across fields ranging from epidemiology~\cite{Aral2018Jun, Bauch2013Oct} and public health~\cite{Centola2013May} to political communication~\cite{Starbird2023Apr, bovet2019influence, Stieglitz2013Dec} and marketing~\cite{Kempe2003Aug, Chen2010Jul, Bakshy2011influencer}. 
These key nodes can determine whether an epidemic will spread or whether a messaging campaign will achieve its intended impact. 

To understand the effect of reconstructing information cascades on social influence analysis, we conduct case studies using data from two microblogging platforms:~Twitter and Bluesky (see Materials and Methods for details). 
For each platform, we construct two types of resharing networks.
The first, referred to as a \textit{naive} network, is constructed directly from API-provided platform data connecting all resharing nodes to the original poster and disregarding any intermediate users in the cascade. 
The second, referred to as a \textit{reconstructed} network, is generated after applying the PDI method as described above.
Specifically, we generate 900 reconstructed networks---100 for each of the nine parameter settings obtained by combining $\gamma \in \{0.25, 0.5, 0.75\}$ and $\alpha \in \{1.1, 2.0, 3.0\}$. 
Note that the connection of the first resharing node is deterministic, as there is only one possible parent (the root).
Therefore, for cascades with only two nodes (the original post and one reshare), no inference is needed. 
These cascades are included in all resharing networks.

\begin{figure}[t]
    \centering
    \includegraphics[width=.8\linewidth]{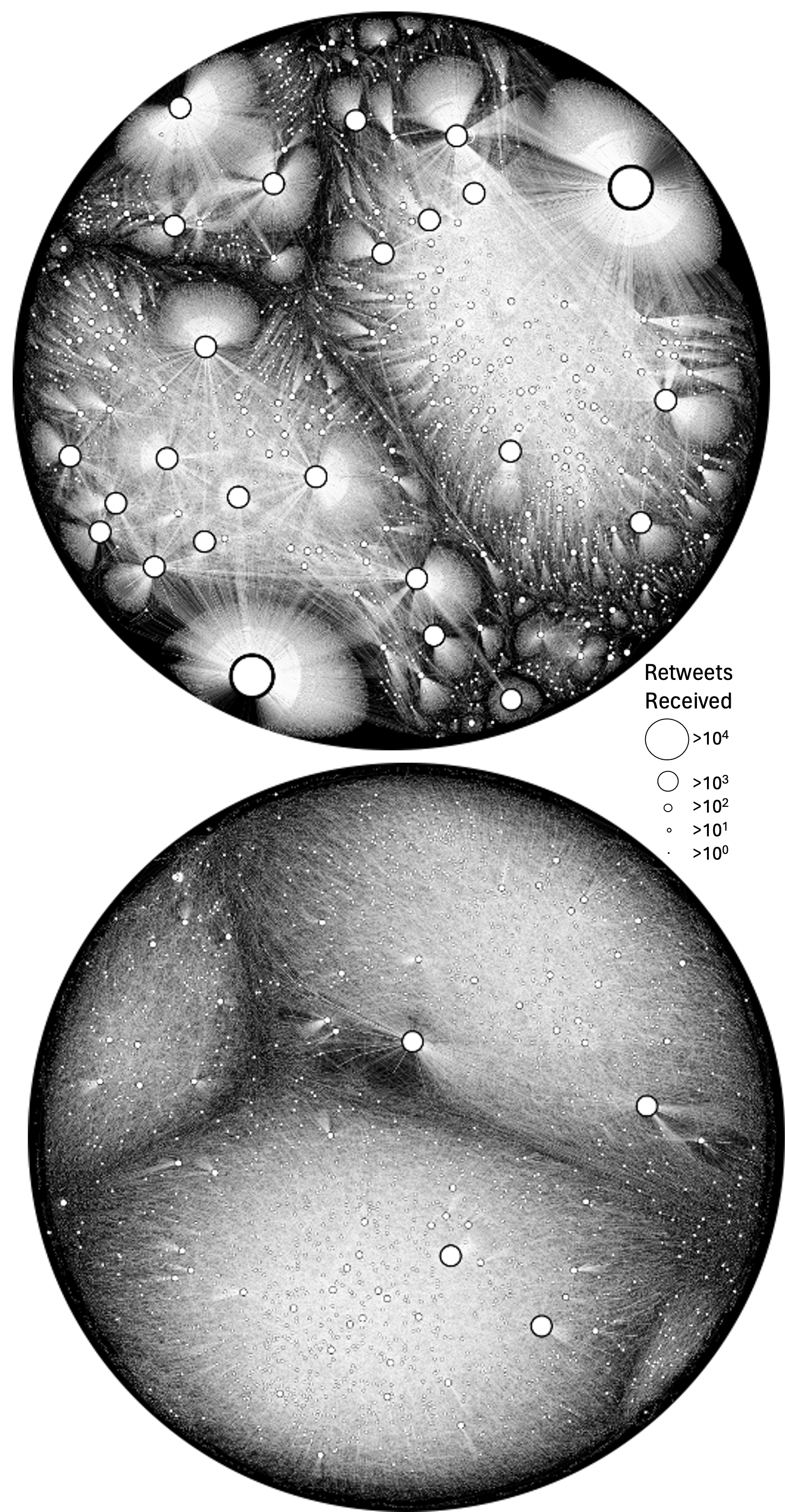}
    \caption{
    Effects of cascade reconstruction on a Twitter resharing network.
    (a) shows the naive network, while (b) displays a version of the same network reconstructed using PDI with parameters $\gamma = 0.5$ and $\alpha = 2.0$. 
    Node size reflects the number of retweets received by an account, with larger nodes representing more influential accounts.
    }
    \label{fig:midterm_reshare_networks}
\end{figure}

Comparing the two networks lets us determine the effects of the reconstruction method on the analysis of node influence. If the results were very similar, it would indicate that the reconstruction process has minimal impact. 
To measure node influence, we calculate node out-strength, or node strength for brevity.
This is a widely recognized and intuitive metric, defined as the total number of reshares a node accumulates~\cite{Jackson2016Mar, Lu2016Sep, Cha2010May}. 
As shown in Figure~\ref{fig:midterm_reshare_networks}, there are important differences in node influence based on the reconstruction method. 
In the naive resharing network, influence is concentrated among a few accounts.
In the reconstructed network, on the other hand, influence is more broadly distributed across many accounts. 

For a more quantitative analysis, our extensive set of reconstructed networks allows us to evaluate both the average impact of the reconstruction process and the robustness of our findings across different parameter settings.
We begin by calculating Spearman's rank correlation $\rho$ between node strength in the naive and reconstructed networks to quantify the changes in relative influence after reconstruction. 
Here, $\rho=1$ signifies that the reconstruction process does not affect relative influence, while lower $\rho$ values indicate that node influence is affected. 

\begin{figure}
    \centering
    \includegraphics[width=\linewidth]{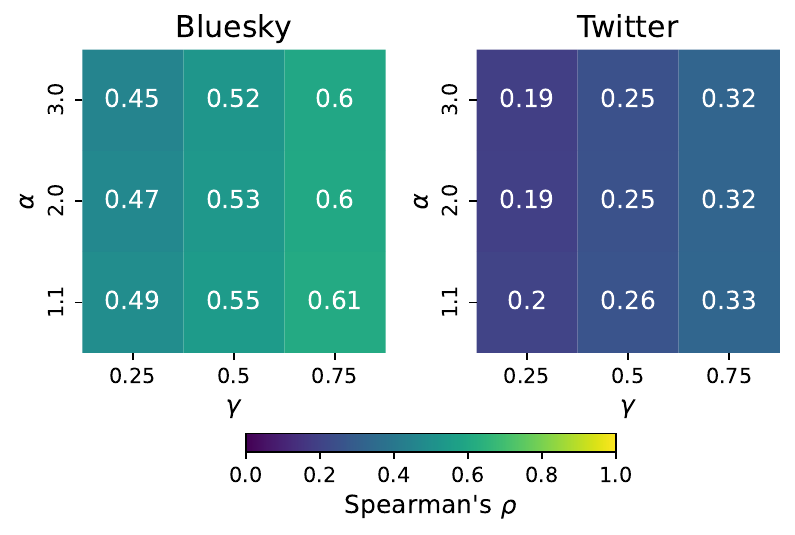}
    \caption{
    Node influence is substantially affected by cascade reconstruction on Bluesky and even more so on Twitter. 
    Heat map cells display the mean Spearman's correlation $\rho$ between node strength values in naive and PDI-reconstructed networks, averaged over 100 versions of the reconstructed network at the specified parameter settings. 
    A $\rho$ value of one means the reconstruction doesn't alter node influence, while values closer to zero suggest significant changes. 
    The maximum standard deviation of correlation values for any parameter setting is 0.001 for Twitter and 0.003 for Bluesky (see the Appendix for full statistics). 
    }
    \label{fig:mean_correlations}
\end{figure}

Figure~\ref{fig:mean_correlations} presents the average correlation values for all tested parameter settings, revealing notable changes in node influence due to the reconstruction process.
In the Bluesky data, $\rho$ values range from 0.45 to 0.61, indicating a moderate shift in influence. 
On Twitter, the $\rho$ values are even lower, between 0.19 and 0.33, pointing to a significant reordering of node influence.
These low correlations highlight the considerable impact of cascade reconstruction on the perceived influence of nodes in both platforms.

\begin{figure*}
    \vspace{-1em}
    \centering
    \includegraphics[width=.5\linewidth]{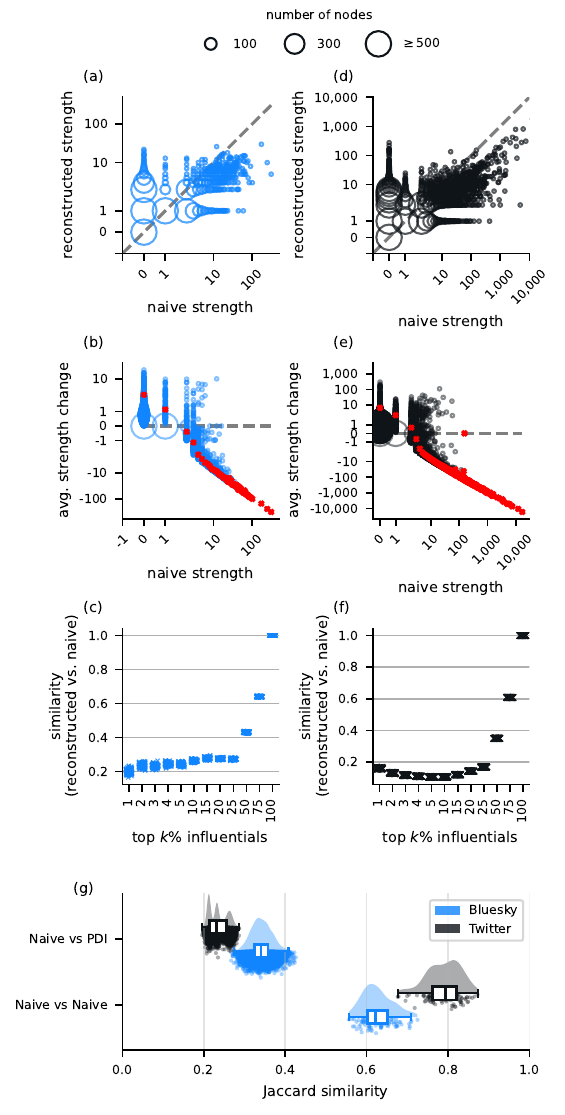}
    \caption{
    Resharing networks reconstructed using the PDI method show substantial shifts in node influence and community detection compared to those built from naive data, on both Bluesky and Twitter.
    Panels (a,~b,~c) present results of the social influence analysis for Bluesky, while panels (d,~e,~f) show results for Twitter; these panels reflect reconstructions using PDI parameters $\gamma = 0.25$ and $\alpha = 3.0$. . 
    Panel (g) provides results in the community detection analysis for both platforms, using all choices of $\gamma$ and $\alpha$. 
    (a,~d): Comparison of node strength between a single version of the PDI-reconstructed network and the corresponding naive network.
    (b,~e): Average change in node strength relative to naive strength, across all 100 PDI reconstructions. 
    Circle sizes in panels (a,~b,~d,~e) represent the number of nodes at each point.
    For visual clarity, we use the same size for all points with 500 or more nodes.
    The red crosses show the median values. 
    (c,~f): Jaccard similarity between the top $k$\% of influential nodes identified based on node strength from reconstructed and naive networks.
    Each marker represents one of the 100 possible comparisons.
    (g): Raincloud distributions of the Jaccard similarity between community partitions from the naive network across 100 Louvain runs (Naive vs.\ Naive) and between the naive network partition and each of the 100 PDI-reconstructed network partitions (Naive vs.\ PDI), across all $\gamma$ and $\alpha$ settings. 
    The upper portion of each distribution shows the probability density, the middle contains box plots (indicating the median, quartiles, and whiskers), and the lower portion displays the individual raw data points.
    }
    \label{fig:inf_nodes}
\end{figure*}

To gain a deeper understanding of how the reconstruction process alters influence, Figure~\ref{fig:inf_nodes}(a-f) presents network changes at a single parameter setting ($\gamma = 0.25$ and $\alpha = 3.0$; we observe similar trends across all parameter settings). 
Panels (a, d) compare node strength between a single PDI-reconstructed network and its corresponding naive network, revealing how influence shifts within the network on both platforms. 
The inclusion of secondary nodes as potential parents rewires network connections, causing some to gain influence while others lose it.

Next, let us examine which nodes gain influence through the reconstruction process, and which ones see it diminish. 
Figure~\ref{fig:inf_nodes}(b, e) shows that, on both platforms, nodes with low strength in the naive network tend to experience a modest increase in strength, while nodes with high initial strength undergo a significant decrease.
Specifically, 56\% of Bluesky users and 91\% of Twitter users exhibit a small average increase in influence after reconstruction, as the influence of secondary users is no longer ignored. 
Only nodes with an initial strength below two on Bluesky and three on Twitter display a median increase in average influence. 
For most nodes with a higher initial strength, the reconstruction process leads to a substantial decrease in average strength. 

Finally, we examine the most influential nodes, defined by their total strength (number of reshares).
For each of the 100 reconstructed networks, we compare the top $k$\% of influential nodes to those in the naive network. 
We measure the overlap between the two sets using the Jaccard index: a value of one indicates that the reconstruction process does not affect the identification of the most influential nodes, while a value of zero signals a substantial influence shift due to the reconstruction process.
This analysis reflects a substantial restructuring of the network, leading to highly dissimilar sets of top influential nodes (Figure~\ref{fig:inf_nodes}(c, f)).
We observe similar patterns in the two use cases: for low values of $k$ (below 25\%), at most we observe a Jaccard similarity of about 33\% between the two network types on Bluesky (panel~(c); $k=15$), while at the lowest, the overlap decreases to around 10\% on Twitter (panel~(f); $k=5$). As more nodes are included in the comparison, the overlap increases as expected.
This result suggests that analyses of information superspreaders based on naive resharing networks may misclassify substantial portions of influential nodes.

\subsection*{Community detection}

Community detection is a central tool in the analysis of online social systems, as communities often reflect coherent social groups, topical interest clusters, or coordinated behavioral patterns~\cite{fortunato2010community}. 
Online communities also affect the viral spread of information in a direct way~\cite{weng2013virality, WengICWSM14}. 
Because community assignments depend entirely on which edges are present and how they are weighted~\cite{newman2004weighted}, any changes to parent–child assignments in cascades directly alter the community partitions produced by detection algorithms---and therefore any downstream analyses, such as measures of polarization~\cite{cinelli2021echo}, that rely on those partitions.

To assess the impact of cascade reconstruction on community structure, we apply the Louvain algorithm~\cite{blondel2008fast} to both naive and PDI-reconstructed resharing networks and measure the similarity between each reconstructed partition and its corresponding naive partition using the Jaccard index~\cite{gates2019clusim, gates2017impact}.
Here a value of one indicates identical community assignments and lower values indicate reconstruction-driven reorganization.
We run the algorithm 100 times on each naive network and once per reconstructed version to separately quantify algorithmic and reconstruction-driven variability.

Figure~\ref{fig:inf_nodes}(g) shows the results.
As a baseline, repeated Louvain runs on the naive networks themselves yield high but imperfect overlap (Bluesky median Jaccard $\approx 60\%$; Twitter median Jaccard $\approx 80\%$), confirming modest variability from the Louvain algorithmic prior to any reconstruction.
Comparisons between PDI-reconstructed and naive partitions show dramatically lower overlap: Twitter median Jaccard $\approx 25\%$ and Bluesky $\approx 30\%$, reductions that are large and statistically robust (Mann-Whitney test $p < 0.001$).
We also observe three distinct peaks in the Twitter Naive vs.\ PDI distribution, likely driven by the choices of $\gamma$; this pattern is absent in Bluesky, possibly reflecting Twitter's larger network size amplifying the structural impact of different reconstruction settings.
Across all scenarios, modularity remains consistently high ($\approx 0.9$), indicating that users form strongly clustered groups based on resharing behavior, largely independent of the specific reconstruction method employed.

Together, these findings show that reliance on naive platform-provided data substantially distorts widely used network-level analyses of both influence and community structure.

\subsection*{Information cascade structure}
\label{sec:vosoughi}

Let us now analyze how decisions made during the reconstruction process affect \textit{individual} information cascades at the microscopic level. 
We posit that if distinct reconstruction methods generate cascades with different structural properties, they will have a substantial impact on downstream analyses.
Given that no platform-provided data exists to validate \textit{any} proposed method, such a finding would raise concerns about the validity of social network studies that rely on network structure.

Based on this premise, we now compare two cascade reconstruction approaches: PDI and Time-Inferred Diffusion (TID), the latter employed in a prominent analysis of true and false rumor cascades on Twitter spanning from 2006 to 2017~\cite{vosoughi2018spread}.
As discussed earlier, TID relies on the assumption that users can only reshare posts from those they follow---a premise we have shown to be increasingly untenable in the era of algorithmically driven feeds (see Cascade reconstruction problems).
Therefore, we explore whether these different reconstruction methods alter the resulting cascades.  

\begin{figure}
    \centering
    \includegraphics[width=\linewidth]{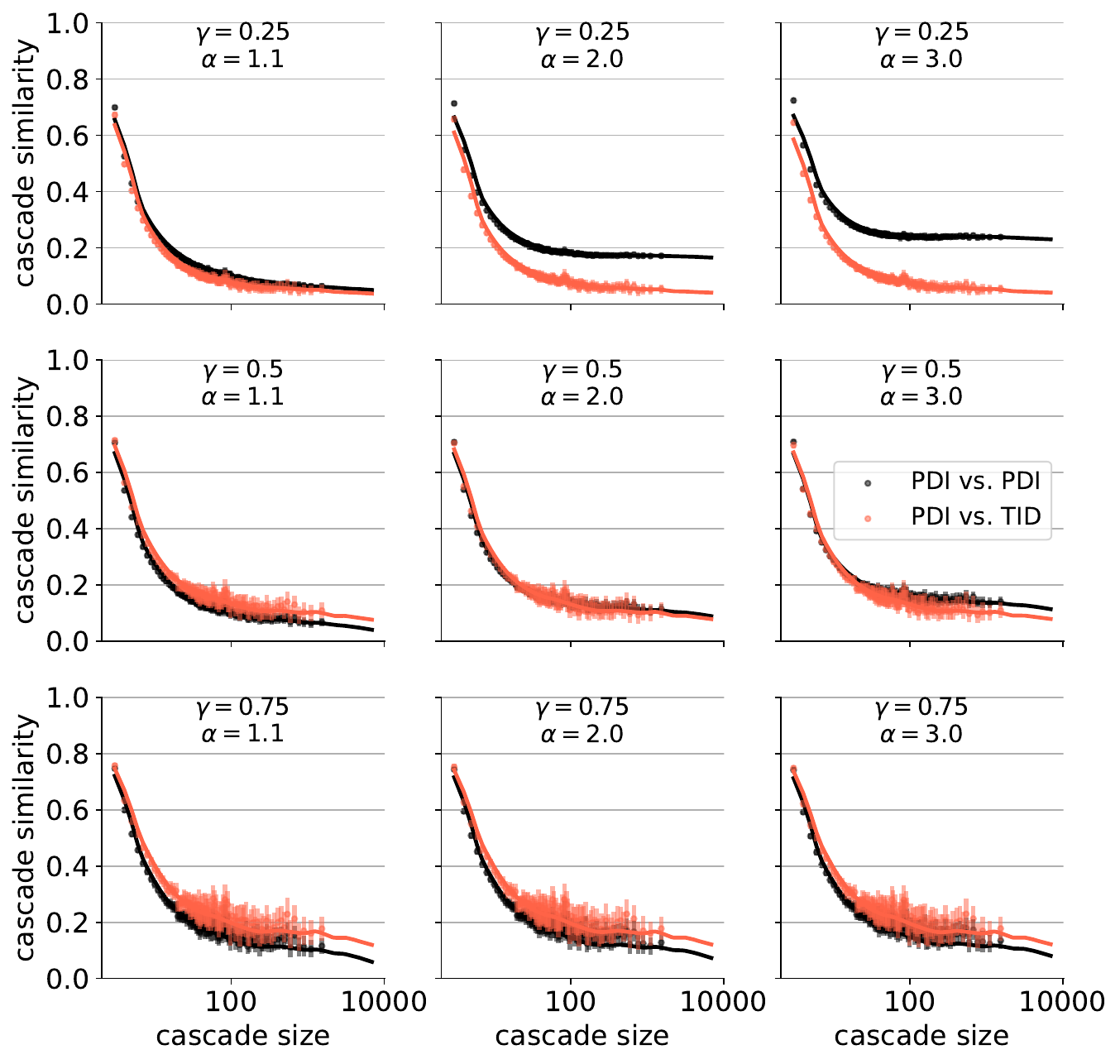}
    \caption{
    Cascades reconstructed in different ways are highly dissimilar, especially larger ones.
    The panels correspond to different PDI reconstruction parameters. 
    Each panel plots the mean cascade similarity, measured using the Jaccard index between edges, as a function of cascade size. We compare PDI-reconstructed cascades to each other and to TID-reconstructed cascades. 
    Points represent the means in 500 equally-sized x-axis bins, while fit lines are generated using locally weighted robust smoothing of the $\sim$28k mean values. Error bars show 95\% confidence intervals calculated from 1,000 bootstraps.
    }
\label{fig:cas_sim}
\end{figure}

Our analysis reconstructs over 40,000 cascades from Vosoughi et al.~\cite{vosoughi2018spread}, originally generated by the TID method, using the PDI approach with the same parameter settings from our earlier analysis: $\gamma \in \{0.25, 0.5, 0.75\}$ and $\alpha \in \{1.1, 2.0, 3.0\}$. 
We focus on cascades with three or more nodes ($n = 28{,}062$), as no inference is required for cascades of size two, where the single resharing user has only one potential parent. 
For each setting, we generate 100 versions of each cascade using PDI and calculate the similarity between the different versions of the same cascade.
We compare the PDI versions of a single cascade against each other ($\binom{100}{2} = 4{,}950$ comparisons) as well as against the TID version (100 comparisons). 
This allows us to study not only how the PDI and TID reconstruction approaches differ from each other, but also the variety of cascades generated by a specific reconstruction heuristic. 
We measure the similarity between two cascades using the Jaccard index of their edge sets. 

Figure~\ref{fig:cas_sim} shows that, on average, different reconstruction heuristics yield highly dissimilar cascades, regardless of PDI parameter settings.
This discrepancy is especially pronounced for larger cascades (size $\gtrsim 100$), with similarity consistently below 0.2 and even dropping below 0.1 when $\gamma = 0.25$.
A similar pattern emerges when comparing different PDI versions against each other.

The above results suggest that reconstruction decisions have a substantial impact on the inferred cascades.
But how do these differences influence the overall topological structure? 
To address this question, let us shift our analysis to the macroscopic level. 
Using all reconstructed cascades from the same dataset, we compare the average distributions of several topological properties based on the 100 cascades produced using each PDI setting as well as those generated with TID. 
We examine three key cascade properties: depth, maximum breadth, and structural virality.
Depth is defined as the longest chain of unique reshares from the original post in the cascade, whereas maximum breadth captures the largest number of users at any single depth in the cascade.
Structural virality is defined as the average shortest-path length between all pairs of nodes in the cascade~\cite{goel2016structural}. 
It estimates the extent to which content spreads through a single, large broadcast (low structural virality) versus multiple levels, where each individual contributes only a small part to the overall spread (high structural virality).

\begin{figure}[t]
    \centering
    \includegraphics[width=.95\linewidth]{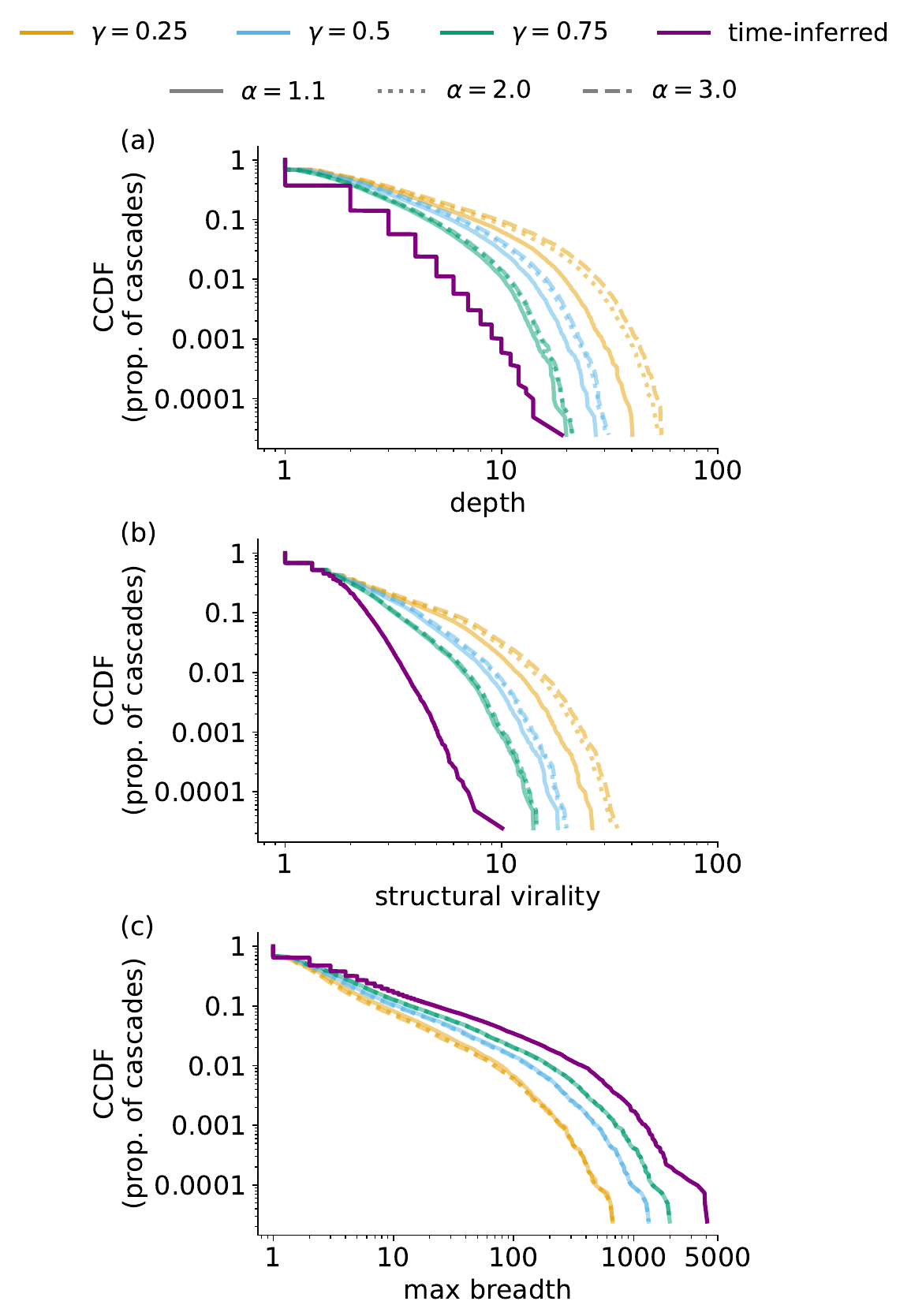}
    \caption{
    The structural properties of cascades are significantly altered by different reconstruction methods.
    Panels show the complementary cumulative distribution functions (CCDF) for (a)~cascade depth, (b)~structural virality, and (c)~maximum breadth. 
    Cascades are reconstructed with the TID (purple) and PDI (other lines) methods. 
    100 versions of each PDI cascade are generated for each parameter setting.
    Lines represent CCDFs based on the mean values across these versions. 
}
    \label{fig:depth_sv_breadth}
\end{figure}

Figure~\ref{fig:depth_sv_breadth} presents the results of this analysis.
For all metrics, we observe that different reconstruction approaches lead to significantly different network distributions, as confirmed by Kolmogorov-Smirnov two-sample tests. 
We have ten reconstruction heuristics---nine PDI settings plus TID---and three metrics, leading to $3 \times \binom{10}{2} = 135$ possible comparisons.  
122 of these (90\%) yield significant differences after applying Bonferroni's correction ($P<0.05$). 
These changes follow expected patterns. 
For instance, as $\gamma$ decreases, giving more weight to the recency of a potential parent's post, both the depth and structural virality of cascades increase. 
Reducing $\gamma$ also lowers the maximum breadth, as the influence of individual prominent accounts with many followers diminishes, and longer chains within a cascade are drawn.
These findings further emphasize how sensitive the inferred network structure is to the specific reconstruction method.

\section*{Discussion}

This study demonstrates that the reconstruction of information cascades can fundamentally reshape diffusion network structures at every level of analysis.
In particular, naive network analyses that rely solely on platform-provided data overestimate the influence of original posters and underestimate the amplification role of intermediate resharers.
Community structure is equally sensitive to cascade inference assumptions, with naive and reconstructed networks exhibiting substantially different community partitions.
Finally, we illustrate how the assumptions embedded within different reconstruction methods significantly affect how we interpret the structure of individual cascades---altering their depth, breadth, and structural virality---and by extension, the collective resharing networks built from them.
Together, these results underscore that reconstruction choices, or the decision to skip reconstruction entirely, shape our understanding of information diffusion at every scale.

These findings were enabled by Probabilistic Diffusion Inference, a novel and flexible approach for reconstructing information cascades. 
This parametric method makes it possible to evaluate the degree to which findings about information diffusion are robust with respect to assumptions made during network reconstruction. 
By combining stochastically generated realizations of each cascade, we are able to construct many versions of weighted resharing (influence) networks. 
This allows us to also explore the variance in outcomes of interest. 

An extension that we have not explored in this manuscript is to probabilistically reflect the influence of multiple parents on a node within a \textit{single} cascade. 
By combining multiple reconstruction realizations, we can represent a cascade as a probability-weighted acyclic graph rather than a simple tree. 
This would make it possible to causally attribute a reshare action not only to one previous action, but to multiple prior exposures~\cite{Dow_Adamic_Friggeri_2013, Friggeri_Adamic_Eckles_Cheng_2014}.

The focus of PDI on the underlying assumptions also makes cascade reconstruction transparent. 
This can help researchers fine-tune assumptions and assess their impact, enabling a deeper exploration of the human and algorithmic factors driving information diffusion. 
For example, how might diffusion dynamics shift if a platform, like X, actively promotes certain political actors, as some have suggested~\cite{Ortutay2024Aug, Murphy2024May, Darcy2024Mar}? 
How does this differ from Meta's new microblogging platform, Threads, which has indicated it will not insert unwanted political content into user feeds~\cite{Fischer2024Feb}? 
Researchers could incorporate node features, like political content, into PDI's probability distributions to explore these and other interesting research questions.

Despite such benefits, PDI should not be considered ``more accurate'' than other techniques, such as the Time-Inferred Diffusion.
The validation of inference methods requires the availability of ground-truth information diffusion data~\cite{SenateHearing2022, Tromble2021Jan, Freelon2018Oct}. 
If platforms were to publicly share cascade data with researchers, the PDI framework could be leveraged to refine assumptions and optimize parameter settings for more accurate modeling. 
However, we note that even platforms have to make assumptions about parent attribution, as users may be exposed to a piece of content in different ways prior to sharing, which are not revealed by their specific sharing action.

The substantial divergence we have found between networks reconstructed with different methods underscores potential risks for researchers who study online phenomena. 
This is especially true when relying on naive networks provided by platforms, which can introduce bias in analyses~\cite{weber_exploring_2021, tromble_we_2017, Gonzalez-Bailon2014Jul} even without reconstruction concerns. 
Given the widespread reliance on platform-provided data and the lack of ground truth for diffusion cascades, researchers must approach information diffusion analyses with caution. 
These issues have far-reaching implications for fields that analyze social media networks, such as conservation science~\cite{Toivonen2019May}, political communication~\cite{Starbird2023Apr, bovet2019influence, Stieglitz2013Dec}, public health~\cite{Centola2013May}, and epidemiology~\cite{bedson2021, sooknanan_2020}. 
Future research should focus on identifying which analyses are most sensitive to reconstruction methods and ensuring their robustness across varying assumptions. 

Computational social science must continue to develop innovative analytical approaches that make transparent assumptions and are robust to rigorous methodological scrutiny~\cite{Elmer2023Dec, Ruths2014Nov, Butts2009Jul}. 
Such progress is crucial for deepening our understanding of complex digital ecosystems and the social dynamics that unfold within them.

\section{Methods}
\label{sec:methods}

\subsection*{Twitter data} 

The English-language retweet cascades used to analyze social influence are derived from the Indiana University 2022 U.S. Midterms Multi-Platform Social Media Dataset~\cite{Aiyappa2023Jun}. 
This dataset captures online conversations about the 2022 midterm elections. 
It was gathered using a snowball sampling approach to collect keywords relevant to the 2022 U.S. Midterm elections.
For this study, we randomly sampled 10,000 cascades 
that originated between November 2, 2022, and November 8, 2022 (Election day), while including retweets up to November 15, 2022 to fully capture their diffusion~\cite{goel2016structural}. 
The resulting dataset contains over 187,443 tweets shared by 128,930 unique users.

\subsection*{Bluesky data}

We collected data from Bluesky between March 1--14, 2024, using the public Firehose endpoint, which streams all posts shared on the platform~\cite{BskyFirehose}. 
We then randomly sampled 5,000 repost cascades originating in the first seven days of this period, following the platform's public launch~\cite{Sahneh2024Aug}. 
The same sampling procedure used for the U.S. Midterm dataset was applied, capturing reposts up to one week later (March 21, 2024). 
We excluded 290 cascades from our analysis: 271 due to missing metadata for at least one user's follower count, and 19 because of timestamp discrepancies caused by Bluesky's distributed architecture~\cite{bsky-timestamps}. 
This resulted in a final dataset of 4,710 cascades consisting of 21,338 posts from 15,550 users.

\subsection*{Twitter rumor cascades data} 

We analyze topological network properties using a dataset of rumor cascades from Twitter~\cite{vosoughi2018spread}, 
provided by the authors in a pre-processed and anonymized format for replication purposes. 
The original study gathered retweet cascades of both true and false content, verified by six independent fact-checking organizations.
Specifically, the authors started from tweets that received English-language replies containing links to fact-checking articles.
The initial dataset included approximately 126,000 English-language rumor cascades shared on Twitter by over 3 million users between 2006 and 2017.
We excluded 84,221 cascades without any retweets.
Additionally, since the PDI method requires follower counts, we also removed 1,242 cascades where this information was missing for at least one user in the cascade.
This resulted in a final dataset of 40,839 cascades for analysis.

\subsection*{Probabilistic Diffusion Inference}

We introduce our Probabilistic Diffusion Inference (PDI) method to estimate the likelihood that each user within a social media cascade is the original source of content for subsequent resharers.
We do not aim to validate this method's absolute efficacy---an evaluation that is infeasible given the absence of ground-truth diffusion data. 
Rather, the goal of this approach is to demonstrate how different modeling assumptions can substantially alter downstream analyses of influence, diffusion structure, and network dynamics.

Consider a cascade $c$ involving a sequence of $N_c$ users, $U^c = \{u^c_0, u^c_1, \dots, u^c_{N_c}\}$,
where $u^c_0$ is the originator of the content, and each subsequent user $u^c_i$ represents the $i$-th person to reshare it. 
To infer the parent of $u^c_i$, i.e., the source of $u^c_i$'s reshare, PDI considers the subset of all prior users $U^c_i = \{ u_j^c \; \forall j < i \}$ as potential parents, each with a probability $p_{ij}$ of being selected as the parent of $u^c_i$.
For all resharing users in the cascade, a potential parent is selected as the parent based on these probabilities. 

PDI enables flexible computation of the probabilities $p_{ij}$ using researcher-defined assumptions. 
In this work, we adopt two common assumptions. 
First, users with more followers are more likely to be the parents of a resharing user~\cite{Myers2014Apr}, which we refer to as the \textit{followers} assumption. 
Second, users who recently reshared the content are more likely to be the true parents of subsequent users~\cite{Crane2008Oct, Notarmuzi2022Mar}, referred to as the \textit{recency} assumption. These assumptions are visually represented in Figure~\ref{fig:reconstruction}(d, e, f).

The probability of a potential parent $u_j \in U_i^c$ according to the followers assumption is given by:
\begin{equation}
    p_{ij}^\mathcal{F} = \frac{F(u_j)}{\sum_{u_k \in  U_i^c} F(u_k)}
\end{equation}
where $F(u)$ represents the mean number of followers of user $u$ during the observed period. 

The recency assumption is modeled using a power-law distribution, which has been shown to describe the timing of resharing behavior on social media platforms~\cite{Crane2008Oct, Notarmuzi2022Mar}:  
\begin{equation}
    P(\Delta_{ij}^c) = \frac{\alpha - 1}{\Delta_{\text{min}}} \left( \frac{\Delta_{ij}^c}{\Delta_{\text{min}}} \right)^{-\alpha},
\end{equation}
where $\Delta_{ij}^c$ is the time (in seconds) between the post by potential parent $u_j^c$ and the reshare by user $u_i^c$, $\Delta_{\text{min}}$ is a minimum time delay (one second), and $\alpha$ is a parameter that expresses the tendency for reshares to be clustered in time, with higher values giving more importance to recent posts. 
Then, the probability of potential parent $u_j$ according to the recency assumption is calculated by:
\begin{equation}
    p_{ij}^\mathcal{T} = \frac{P(\Delta_{ij}^c)}{\sum_{u_k \in  U_i^c} P(\Delta_{ik}^c)}.
\end{equation}

We consider the followers and recency assumptions as independent factors and combine them using a weighting parameter $\gamma$ ($0 \leq \gamma \leq 1$), yielding the overall probability that $u_j^c$ is the true parent of $u_i^c$:
\begin{equation}
    p_{ij} = \gamma p_{ij}^\mathcal{F} + (1 - \gamma) p_{ij}^\mathcal{T}.
\end{equation}

In sum, we calculate two probabilities for each potential parent node: one based on their number of followers and the other taking into account the recency of their activity.
The relative influence of these two factors is adjusted using a parameter $\gamma$: higher values give more weight to follower counts, while lower values prioritize reshare recency.

\subsection*{Community detection}

We perform community detection using the Louvain algorithm~\cite{blondel2008fast} as implemented in the \textit{igraph} library, treating all networks as undirected and preserving edge weights.
For each naive network, we compute a reference partition, then run the algorithm an additional 100 times comparing each run to that reference. For each reconstructed network version, we run the algorithm 100 times, comparing each run to the naive reference partition.
Clustering similarity is computed using the Jaccard index as defined in the \textit{clusim} library~\cite{gates2019clusim, gates2017impact}.

\section*{Funding statement}

This work was supported in part by the Knight Foundation, Craig Newmark Philanthropies, the Italian Ministry of Education (PRIN PNRR grant CODE prot. P2022AKRZ9 and PRIN grant DEMON prot. 2022BAXSPY), and DARPA (grant W911NF-17-C-0094).

\section*{Code and data}

Code and data are available in public repositories on GitHub (\href{https://github.com/osome-iu/cascade_reconstruction}{github.com/osome-iu/cascade\_reconstruction}) and Zenodo (\href{https://doi.org/10.5281/zenodo.13994029}{doi.org/10.5281/zenodo.13994029}).

This study, focusing on public data, poses minimal risk to human subjects. 
Consequently, the Indiana University Institutional Review Board has exempted it from review (protocol numbers 1102004860 and 23757).

\section*{Acknowledgments}

We are grateful to Sandra Gonzáles-Bailón, Alessandro Flammini, Jonas Juul, and Dean Eckles for their valuable feedback.
We also thank Soroush Vosoughi, Deb Roy, and Sinan Aral for generously providing the data used in our information cascade structure analyses. 
This work used JetStream2~\cite{Hancock2021Jetstream2, BoernerJetstream2023} at Indiana University through allocation CIS200033 from the Advanced Cyberinfrastructure Coordination Ecosystem: Services \& Support (ACCESS) program, which is supported by National Science Foundation grants 2138259, 2138286, 2138307, 2137603, and 2138296.
We also acknowledge the team of developers at the Observatory on Social Media for their assistance in collecting Bluesky data.

\section*{Author contributions}

MRD and FP conceptualized the study design.
FP, MRD and RA collected and curated data. 
Code was primarily written by MRD with review and input from FP.
MRD and FP conducted the analyses.
MRD, FP, and DP created the visualizations.
The Probabilistic Diffusion Inference method was developed by MRD with input from RA and JB.
The first draft was written by MRD and FP, with input from all authors. 
FM provided guidance throughout the study.
Funding was acquired by FM and FP.

\bibliography{main.bib}

\clearpage
\setcounter{page}{1}
\appendix

\counterwithin{figure}{section}
\counterwithin{table}{section}

\counterwithout{figure}{section}
\counterwithout{table}{section}

\renewcommand{\thefigure}{A\arabic{figure}}
\renewcommand{\thetable}{A\arabic{table}}

\setcounter{figure}{0}
\setcounter{table}{0}

\renewcommand{\thesection}{\arabic{section}}

\renewcommand{\thesubsection}{\thesection.\arabic{subsection}}

\section*{Appendix}

\subsection*{Bibliographic analysis}

\begin{figure*}[h]
    \centering
    \includegraphics[width=\linewidth]{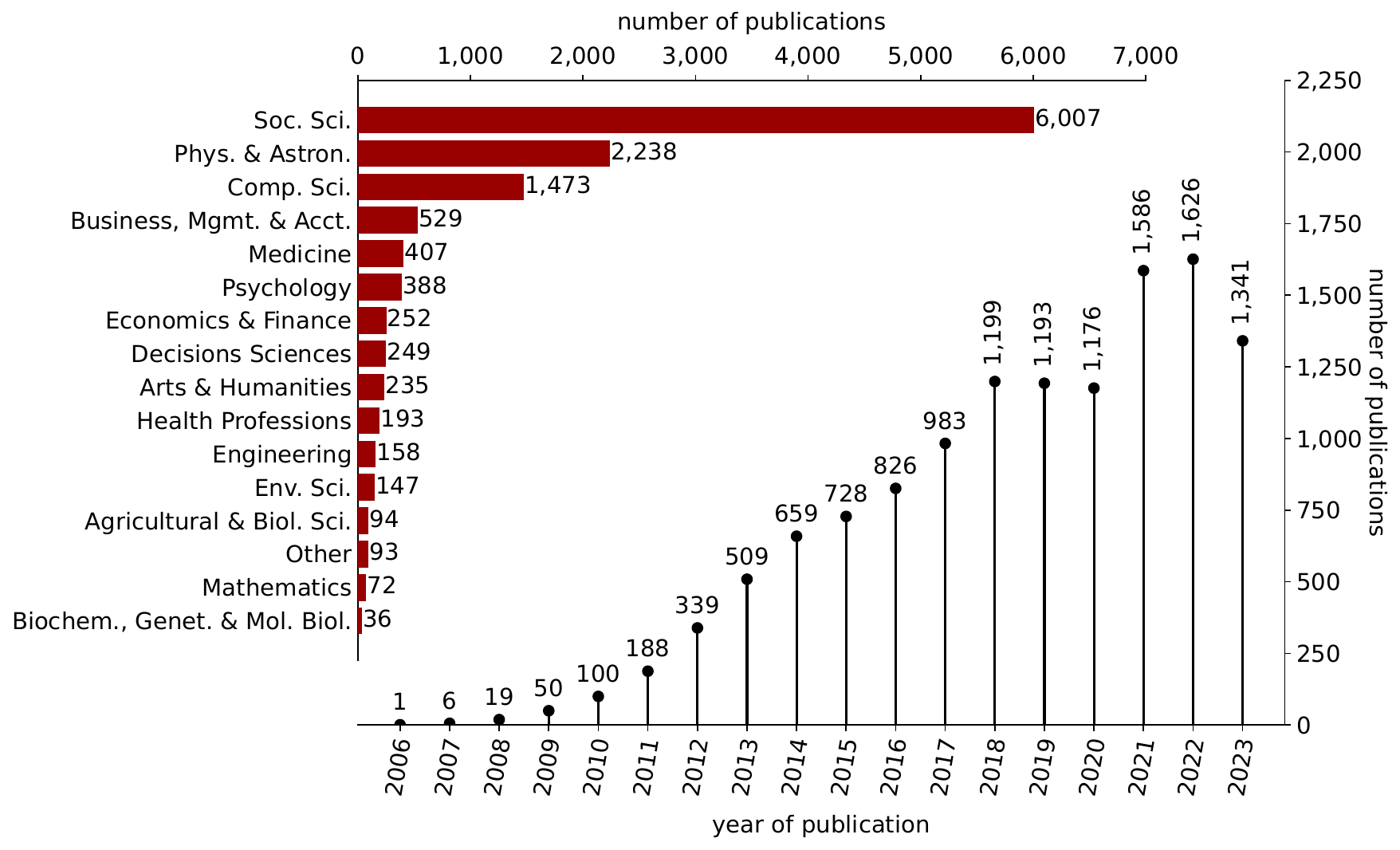}
    \caption{
    Research on information diffusion and social media has grown rapidly since the early 2000s across various fields.
    The barplot in the top left panel displays the cumulative number of peer-reviewed publications across various academic fields, from 2006 to 2023.
    The time series in the bottom right panel breaks down publication trends annually over the same period.
    }
    \label{fig:lit-review}
\end{figure*}

We analyze bibliographic data from OpenAlex~\cite{Priem2022OpenAlex} to track growing interest in social media information diffusion, shown in Figure~\ref{fig:lit-review}. 
To collect this data we query the \texttt{search-works} endpoint of the OpenAlex Application Programming Interface (API).\footnote{\url{https://docs.openalex.org/how-to-use-the-api/api-overview}}
We employ boolean search parameters to match a wide range of publications that are clearly related to information diffusion on popular social media platforms.
We employ the search query
\begin{quote}
    (``information diffusion'' OR ``diffusion of information'' OR ``information spread'' OR ``spread of information'') AND (``social media'' OR ``facebook'' OR ``twitter'' OR ``reddit'')
\end{quote}
to return entities that find exact matches (case insensitive) within titles, abstracts, or full text.
Our search was not limited by time, retrieving all works in the database that matched our query.
Despite the fact that only a subset of the OpenAlex database contains full text, we obtained 19,294 matching works.
For our analysis, we narrowed the focus to peer-reviewed articles and conference publications from 2006 onward, the year Facebook opened to the public.\footnote{\url{https://www.facebook.com/notes/262051265158581/}}
This filtering removed 6,723 publications from the dataset initially returned by our query.

The ten fields with the smallest number of publications that were grouped together in Figure~\ref{fig:lit-review} to create the ``Other'' group are: ``Neuroscience'' ($n=33$); ``Earth and Planetary Sciences'' ($n=16$); ``Immunology and Microbiology'' ($n=15$); ``Dentistry'' ($n=14$); ``Pharmacology, Toxicology and Pharmaceutics'' ($n=5$); ``Energy'' ($n=4$); ``Nursing'' ($n=3$); ``Materials Science'' ($n=1$); ``Chemistry'' ($n=1$); ``Veterinary'' ($n=1$).
Collectively, these papers account for 0.74\% of the publications in our collection.

\subsection*{Node strength correlations}

Table~\ref{tab:si:meancorr} presents the statistics for the mean and standard deviation of node strength correlation values between naive and reconstructed networks.
Correlations are calculated using Spearman's $\rho$.

\begin{table*}[h]
    \centering
    \begin{tabular}{cc cc cc}
        &  & \multicolumn{2}{c}{Twitter} & \multicolumn{2}{c}{Bluesky} \\
        \cmidrule(lr){3-4} \cmidrule(lr){5-6}
        $\gamma$ & $\alpha$ & $\bar{\rho}$ & $\sigma$ & $\bar{\rho}$ & $\sigma$ \\
        \midrule
        0.25 & 1.1  & 0.201 & 0.001 & 0.488 & 0.003 \\
        0.25 & 2.0 & 0.190 & 0.001 & 0.466 & 0.003 \\
        0.25 & 3.0 & 0.187 & 0.001 & 0.451 & 0.003 \\
   \midrule
        0.5 & 1.1 & 0.259 & 0.001 & 0.547 & 0.003 \\
        0.5 & 2.0 & 0.252 & 0.001 & 0.533 & 0.003 \\
        0.5 & 3.0 & 0.250 & 0.001 & 0.525 & 0.003 \\
        \midrule
        0.75 & 1.1 & 0.327 & 0.001 & 0.606 & 0.003 \\
        0.75 & 2.0 & 0.323 & 0.001 & 0.599 & 0.003 \\
        0.75 & 3.0 & 0.322 & 0.001 & 0.595 & 0.003 \\
        \bottomrule
    \end{tabular}
    \caption{Mean and standard deviation of Spearman's correlations between node out-strengths of naive and reconstructed networks.}
    \label{tab:si:meancorr}
\end{table*}

\subsection*{Comparing cascade metric distributions}

In the \textit{Information cascade structure} section of the main text, we calculate the depth, breadth, and structural virality of cascades and compare the distributions of these metrics across different reconstruction approaches. 
To determine whether these distributions differ significantly, we perform Kolmogorov-Smirnov two-sample tests for all possible comparisons. 
The results for depth, maximum breadth, and structural virality are presented in Tables~\ref{tab:ks_results_depth}, \ref{tab:ks_results_breadth}, and \ref{tab:ks_results_sv}, respectively. 
To account for multiple comparisons, we apply a Bonferroni correction across 45 comparisons, treating each metric as its own family of comparisons.
Out of 135 possible comparisons, 122 (90\%) are significant ($P<0.05$). We omit tables due to space limitations.

\begin{table*}
\centering
\caption{Kolmogorov-Smirnoff statistics for comparing depth distributions. Rows containing ``TID'' represent comparisons to distributions based on the Time-Inferred Diffusion method. All values are rounded to two decimal points.}
\label{tab:ks_results_depth}
\begin{tabular}{ccccccccc}
\toprule
\# & $\gamma_1$ & $\alpha_1$ & $\gamma_2$ & $\alpha_2$ & statistic & $P$ & $P$ adj.$^\dagger$ & Sig. \\
\midrule
1 & 0.25 & 1.10 & 0.25 & 2.00 & 0.02 & 0.00 & 0.00 & *** \\
2 & 0.25 & 1.10 & 0.25 & 3.00 & 0.04 & 0.00 & 0.00 & *** \\
3 & 0.25 & 1.10 & 0.50 & 1.10 & 0.05 & 0.00 & 0.00 & *** \\
4 & 0.25 & 1.10 & 0.50 & 2.00 & 0.04 & 0.00 & 0.00 & *** \\
5 & 0.25 & 1.10 & 0.50 & 3.00 & 0.03 & 0.00 & 0.00 & *** \\
6 & 0.25 & 1.10 & 0.75 & 1.10 & 0.12 & 0.00 & 0.00 & *** \\
7 & 0.25 & 1.10 & 0.75 & 2.00 & 0.11 & 0.00 & 0.00 & *** \\
8 & 0.25 & 1.10 & 0.75 & 3.00 & 0.10 & 0.00 & 0.00 & *** \\
9 & 0.25 & 1.10 & TID& TID& 0.35 & 0.00 & 0.00 & *** \\
10 & 0.25 & 2.00 & 0.25 & 3.00 & 0.02 & 0.00 & 0.00 & ** \\
11 & 0.25 & 2.00 & 0.50 & 1.10 & 0.08 & 0.00 & 0.00 & *** \\
12 & 0.25 & 2.00 & 0.50 & 2.00 & 0.06 & 0.00 & 0.00 & *** \\
13 & 0.25 & 2.00 & 0.50 & 3.00 & 0.05 & 0.00 & 0.00 & *** \\
14 & 0.25 & 2.00 & 0.75 & 1.10 & 0.14 & 0.00 & 0.00 & *** \\
15 & 0.25 & 2.00 & 0.75 & 2.00 & 0.13 & 0.00 & 0.00 & *** \\
16 & 0.25 & 2.00 & 0.75 & 3.00 & 0.12 & 0.00 & 0.00 & *** \\
17 & 0.25 & 2.00 & TID& TID& 0.36 & 0.00 & 0.00 & *** \\
18 & 0.25 & 3.00 & 0.50 & 1.10 & 0.09 & 0.00 & 0.00 & *** \\
19 & 0.25 & 3.00 & 0.50 & 2.00 & 0.07 & 0.00 & 0.00 & *** \\
20 & 0.25 & 3.00 & 0.50 & 3.00 & 0.06 & 0.00 & 0.00 & *** \\
21 & 0.25 & 3.00 & 0.75 & 1.10 & 0.15 & 0.00 & 0.00 & *** \\
22 & 0.25 & 3.00 & 0.75 & 2.00 & 0.14 & 0.00 & 0.00 & *** \\
23 & 0.25 & 3.00 & 0.75 & 3.00 & 0.14 & 0.00 & 0.00 & *** \\
24 & 0.25 & 3.00 & TID& TID& 0.37 & 0.00 & 0.00 & *** \\
25 & 0.50 & 1.10 & 0.50 & 2.00 & 0.02 & 0.00 & 0.00 & *** \\
26 & 0.50 & 1.10 & 0.50 & 3.00 & 0.03 & 0.00 & 0.00 & *** \\
27 & 0.50 & 1.10 & 0.75 & 1.10 & 0.07 & 0.00 & 0.00 & *** \\
28 & 0.50 & 1.10 & 0.75 & 2.00 & 0.06 & 0.00 & 0.00 & *** \\
29 & 0.50 & 1.10 & 0.75 & 3.00 & 0.05 & 0.00 & 0.00 & *** \\
30 & 0.50 & 1.10 & TID& TID& 0.31 & 0.00 & 0.00 & *** \\
31 & 0.50 & 2.00 & 0.50 & 3.00 & 0.01 & 0.00 & 0.18 &  \\
32 & 0.50 & 2.00 & 0.75 & 1.10 & 0.08 & 0.00 & 0.00 & *** \\
33 & 0.50 & 2.00 & 0.75 & 2.00 & 0.07 & 0.00 & 0.00 & *** \\
34 & 0.50 & 2.00 & 0.75 & 3.00 & 0.07 & 0.00 & 0.00 & *** \\
35 & 0.50 & 2.00 & TID& TID& 0.32 & 0.00 & 0.00 & *** \\
36 & 0.50 & 3.00 & 0.75 & 1.10 & 0.09 & 0.00 & 0.00 & *** \\
37 & 0.50 & 3.00 & 0.75 & 2.00 & 0.09 & 0.00 & 0.00 & *** \\
38 & 0.50 & 3.00 & 0.75 & 3.00 & 0.08 & 0.00 & 0.00 & *** \\
39 & 0.50 & 3.00 & TID& TID& 0.33 & 0.00 & 0.00 & *** \\
40 & 0.75 & 1.10 & 0.75 & 2.00 & 0.01 & 0.03 & 1.00 &  \\
41 & 0.75 & 1.10 & 0.75 & 3.00 & 0.02 & 0.00 & 0.00 & ** \\
42 & 0.75 & 1.10 & TID& TID& 0.31 & 0.00 & 0.00 & *** \\
43 & 0.75 & 2.00 & 0.75 & 3.00 & 0.01 & 0.19 & 1.00 &  \\
44 & 0.75 & 2.00 & TID& TID& 0.31 & 0.00 & 0.00 & *** \\
45 & 0.75 & 3.00 & TID& TID& 0.31 & 0.00 & 0.00 & *** \\
\midrule
\multicolumn{8}{l}{Significance codes: *** $P< 0.001$, ** $P< 0.01$, * $P< 0.05$} \\
\multicolumn{8}{l}{$\dagger$ Using Bonferroni's method with 45 comparisons} \\
\bottomrule
\end{tabular}
\end{table*}

\begin{table*}
\centering
\caption{Kolmogorov-Smirnoff statistics for comparing maximum breadth distributions. Rows containing ``TID'' represent comparisons to distributions based on the Time-Inferred Diffusion method. All values are rounded to two decimal points.}
\label{tab:ks_results_breadth}
\begin{tabular}{ccccccccc}
\toprule
\# & $\gamma_1$ & $\alpha_1$ & $\gamma_2$ & $\alpha_2$ & statistic & $P$ & $P$ adj.$^\dagger$ & Sig. \\
\midrule
1 & 0.25 & 1.10 & 0.25 & 2.00 & 0.02 & 0.00 & 0.00 & *** \\
2 & 0.25 & 1.10 & 0.25 & 3.00 & 0.04 & 0.00 & 0.00 & *** \\
3 & 0.25 & 1.10 & 0.50 & 1.10 & 0.05 & 0.00 & 0.00 & *** \\
4 & 0.25 & 1.10 & 0.50 & 2.00 & 0.04 & 0.00 & 0.00 & *** \\
5 & 0.25 & 1.10 & 0.50 & 3.00 & 0.04 & 0.00 & 0.00 & *** \\
6 & 0.25 & 1.10 & 0.75 & 1.10 & 0.10 & 0.00 & 0.00 & *** \\
7 & 0.25 & 1.10 & 0.75 & 2.00 & 0.09 & 0.00 & 0.00 & *** \\
8 & 0.25 & 1.10 & 0.75 & 3.00 & 0.09 & 0.00 & 0.00 & *** \\
9 & 0.25 & 1.10 & TID& TID& 0.20 & 0.00 & 0.00 & *** \\
10 & 0.25 & 2.00 & 0.25 & 3.00 & 0.02 & 0.00 & 0.00 & *** \\
11 & 0.25 & 2.00 & 0.50 & 1.10 & 0.07 & 0.00 & 0.00 & *** \\
12 & 0.25 & 2.00 & 0.50 & 2.00 & 0.06 & 0.00 & 0.00 & *** \\
13 & 0.25 & 2.00 & 0.50 & 3.00 & 0.06 & 0.00 & 0.00 & *** \\
14 & 0.25 & 2.00 & 0.75 & 1.10 & 0.12 & 0.00 & 0.00 & *** \\
15 & 0.25 & 2.00 & 0.75 & 2.00 & 0.12 & 0.00 & 0.00 & *** \\
16 & 0.25 & 2.00 & 0.75 & 3.00 & 0.11 & 0.00 & 0.00 & *** \\
17 & 0.25 & 2.00 & TID& TID& 0.22 & 0.00 & 0.00 & *** \\
18 & 0.25 & 3.00 & 0.50 & 1.10 & 0.09 & 0.00 & 0.00 & *** \\
19 & 0.25 & 3.00 & 0.50 & 2.00 & 0.08 & 0.00 & 0.00 & *** \\
20 & 0.25 & 3.00 & 0.50 & 3.00 & 0.07 & 0.00 & 0.00 & *** \\
21 & 0.25 & 3.00 & 0.75 & 1.10 & 0.13 & 0.00 & 0.00 & *** \\
22 & 0.25 & 3.00 & 0.75 & 2.00 & 0.13 & 0.00 & 0.00 & *** \\
23 & 0.25 & 3.00 & 0.75 & 3.00 & 0.13 & 0.00 & 0.00 & *** \\
24 & 0.25 & 3.00 & TID& TID& 0.23 & 0.00 & 0.00 & *** \\
25 & 0.50 & 1.10 & 0.50 & 2.00 & 0.01 & 0.00 & 0.02 & * \\
26 & 0.50 & 1.10 & 0.50 & 3.00 & 0.02 & 0.00 & 0.00 & *** \\
27 & 0.50 & 1.10 & 0.75 & 1.10 & 0.05 & 0.00 & 0.00 & *** \\
28 & 0.50 & 1.10 & 0.75 & 2.00 & 0.05 & 0.00 & 0.00 & *** \\
29 & 0.50 & 1.10 & 0.75 & 3.00 & 0.04 & 0.00 & 0.00 & *** \\
30 & 0.50 & 1.10 & TID& TID& 0.16 & 0.00 & 0.00 & *** \\
31 & 0.50 & 2.00 & 0.50 & 3.00 & 0.01 & 0.03 & 1.00 &  \\
32 & 0.50 & 2.00 & 0.75 & 1.10 & 0.06 & 0.00 & 0.00 & *** \\
33 & 0.50 & 2.00 & 0.75 & 2.00 & 0.05 & 0.00 & 0.00 & *** \\
34 & 0.50 & 2.00 & 0.75 & 3.00 & 0.05 & 0.00 & 0.00 & *** \\
35 & 0.50 & 2.00 & TID& TID& 0.17 & 0.00 & 0.00 & *** \\
36 & 0.50 & 3.00 & 0.75 & 1.10 & 0.06 & 0.00 & 0.00 & *** \\
37 & 0.50 & 3.00 & 0.75 & 2.00 & 0.06 & 0.00 & 0.00 & *** \\
38 & 0.50 & 3.00 & 0.75 & 3.00 & 0.06 & 0.00 & 0.00 & *** \\
39 & 0.50 & 3.00 & TID& TID& 0.18 & 0.00 & 0.00 & *** \\
40 & 0.75 & 1.10 & 0.75 & 2.00 & 0.01 & 0.14 & 1.00 &  \\
41 & 0.75 & 1.10 & 0.75 & 3.00 & 0.01 & 0.00 & 0.08 &  \\
42 & 0.75 & 1.10 & TID& TID& 0.15 & 0.00 & 0.00 & *** \\
43 & 0.75 & 2.00 & 0.75 & 3.00 & 0.01 & 0.47 & 1.00 &  \\
44 & 0.75 & 2.00 & TID& TID& 0.15 & 0.00 & 0.00 & *** \\
45 & 0.75 & 3.00 & TID& TID& 0.15 & 0.00 & 0.00 & *** \\
\midrule
\multicolumn{8}{l}{Significance codes: *** $P< 0.001$, ** $P< 0.01$, * $P< 0.05$} \\
\multicolumn{8}{l}{$\dagger$ Using Bonferroni's method with 45 comparisons} \\
\bottomrule
\end{tabular}
\end{table*}

\begin{table*}
\centering
\caption{Kolmogorov-Smirnoff statistics for comparing structural virality distributions. Rows containing ``TID'' represent comparisons to distributions based on the Time-Inferred Diffusion method. All values are rounded to two decimal points.}
\label{tab:ks_results_sv}
\begin{tabular}{ccccccccc}
\toprule
\# & $\gamma_1$ & $\alpha_1$ & $\gamma_2$ & $\alpha_2$ & statistic & $P$ & $P$ adj.$^\dagger$ & Sig. \\
\midrule
1 & 0.25 & 1.10 & 0.25 & 2.00 & 0.01 & 0.00 & 0.01 & * \\
2 & 0.25 & 1.10 & 0.25 & 3.00 & 0.02 & 0.00 & 0.00 & *** \\
3 & 0.25 & 1.10 & 0.50 & 1.10 & 0.05 & 0.00 & 0.00 & *** \\
4 & 0.25 & 1.10 & 0.50 & 2.00 & 0.04 & 0.00 & 0.00 & *** \\
5 & 0.25 & 1.10 & 0.50 & 3.00 & 0.04 & 0.00 & 0.00 & *** \\
6 & 0.25 & 1.10 & 0.75 & 1.10 & 0.09 & 0.00 & 0.00 & *** \\
7 & 0.25 & 1.10 & 0.75 & 2.00 & 0.08 & 0.00 & 0.00 & *** \\
8 & 0.25 & 1.10 & 0.75 & 3.00 & 0.08 & 0.00 & 0.00 & *** \\
9 & 0.25 & 1.10 & TID& TID& 0.18 & 0.00 & 0.00 & *** \\
10 & 0.25 & 2.00 & 0.25 & 3.00 & 0.01 & 0.02 & 0.89 &  \\
11 & 0.25 & 2.00 & 0.50 & 1.10 & 0.06 & 0.00 & 0.00 & *** \\
12 & 0.25 & 2.00 & 0.50 & 2.00 & 0.05 & 0.00 & 0.00 & *** \\
13 & 0.25 & 2.00 & 0.50 & 3.00 & 0.05 & 0.00 & 0.00 & *** \\
14 & 0.25 & 2.00 & 0.75 & 1.10 & 0.10 & 0.00 & 0.00 & *** \\
15 & 0.25 & 2.00 & 0.75 & 2.00 & 0.09 & 0.00 & 0.00 & *** \\
16 & 0.25 & 2.00 & 0.75 & 3.00 & 0.09 & 0.00 & 0.00 & *** \\
17 & 0.25 & 2.00 & TID& TID& 0.19 & 0.00 & 0.00 & *** \\
18 & 0.25 & 3.00 & 0.50 & 1.10 & 0.07 & 0.00 & 0.00 & *** \\
19 & 0.25 & 3.00 & 0.50 & 2.00 & 0.06 & 0.00 & 0.00 & *** \\
20 & 0.25 & 3.00 & 0.50 & 3.00 & 0.06 & 0.00 & 0.00 & *** \\
21 & 0.25 & 3.00 & 0.75 & 1.10 & 0.10 & 0.00 & 0.00 & *** \\
22 & 0.25 & 3.00 & 0.75 & 2.00 & 0.10 & 0.00 & 0.00 & *** \\
23 & 0.25 & 3.00 & 0.75 & 3.00 & 0.10 & 0.00 & 0.00 & *** \\
24 & 0.25 & 3.00 & TID& TID& 0.20 & 0.00 & 0.00 & *** \\
25 & 0.50 & 1.10 & 0.50 & 2.00 & 0.01 & 0.01 & 0.54 &  \\
26 & 0.50 & 1.10 & 0.50 & 3.00 & 0.02 & 0.00 & 0.00 & *** \\
27 & 0.50 & 1.10 & 0.75 & 1.10 & 0.06 & 0.00 & 0.00 & *** \\
28 & 0.50 & 1.10 & 0.75 & 2.00 & 0.05 & 0.00 & 0.00 & *** \\
29 & 0.50 & 1.10 & 0.75 & 3.00 & 0.05 & 0.00 & 0.00 & *** \\
30 & 0.50 & 1.10 & TID& TID& 0.16 & 0.00 & 0.00 & *** \\
31 & 0.50 & 2.00 & 0.50 & 3.00 & 0.01 & 0.19 & 1.00 &  \\
32 & 0.50 & 2.00 & 0.75 & 1.10 & 0.06 & 0.00 & 0.00 & *** \\
33 & 0.50 & 2.00 & 0.75 & 2.00 & 0.06 & 0.00 & 0.00 & *** \\
34 & 0.50 & 2.00 & 0.75 & 3.00 & 0.06 & 0.00 & 0.00 & *** \\
35 & 0.50 & 2.00 & TID& TID& 0.17 & 0.00 & 0.00 & *** \\
36 & 0.50 & 3.00 & 0.75 & 1.10 & 0.07 & 0.00 & 0.00 & *** \\
37 & 0.50 & 3.00 & 0.75 & 2.00 & 0.07 & 0.00 & 0.00 & *** \\
38 & 0.50 & 3.00 & 0.75 & 3.00 & 0.06 & 0.00 & 0.00 & *** \\
39 & 0.50 & 3.00 & TID& TID& 0.17 & 0.00 & 0.00 & *** \\
40 & 0.75 & 1.10 & 0.75 & 2.00 & 0.01 & 0.64 & 1.00 &  \\
41 & 0.75 & 1.10 & 0.75 & 3.00 & 0.01 & 0.16 & 1.00 &  \\
42 & 0.75 & 1.10 & TID& TID& 0.16 & 0.00 & 0.00 & *** \\
43 & 0.75 & 2.00 & 0.75 & 3.00 & 0.00 & 0.95 & 1.00 &  \\
44 & 0.75 & 2.00 & TID& TID& 0.16 & 0.00 & 0.00 & *** \\
45 & 0.75 & 3.00 & TID& TID& 0.16 & 0.00 & 0.00 & *** \\
\midrule
\multicolumn{8}{l}{Significance codes: *** $P< 0.001$, ** $P< 0.01$, * $P< 0.05$} \\
\multicolumn{8}{l}{$\dagger$ Using Bonferroni's method with 45 comparisons} \\
\bottomrule
\end{tabular}
\end{table*}

\end{document}